# Investigating Artificial Intelligence Digital Sovereignty in Mobile Shopping Apps: A Case Study of Nigeria

*Full Paper*[1]

**George Grispos**
University of Nebraska-Omaha
ggrispos@unomaha.edu

**Sajda Qureshi**
University of Nebraska-Omaha
squreshi@unomaha.edu

## Abstract

The use of e-commerce mobile applications is expanding in Nigeria, creating both opportunities and risks, including fraud and reduced user control over digital technologies, raising concerns about digital sovereignty. This research examines how Artificial Intelligence (AI) in Nigerian mobile applications affects digital sovereignty, examined through platform transparency as a key indicator of user awareness and control. Using an interpretive approach, the research combines the forensic analysis of selected Android applications with contextual document analysis to identify AI features and evaluate disclosure practices. The findings show that AI is widely implemented in the applications, yet transparency about its use remains limited. A socio-economic analysis of Nigeria further shows an increasing dependence on consumer digital platforms, moderate AI awareness, and uneven patterns of interaction. By providing empirical evidence on AI transparency and platform practices, this study advances understanding of individual digital sovereignty and highlights challenges for protecting user control in AI-driven digital environments.

### Keywords

Digital sovereignty, artificial intelligence, mobile applications, trust, transparency.

## Introduction

The rapid expansion of mobile applications and services has transformed everyday life. This transformation is particularly visible in Africa, where mobile connectivity continues to expand and is projected to reach approximately 915 million mobile subscribers by 2030, representing 53% of the population (African Wireless Communications, 2025). Mobile technologies and services already contribute an estimated 7.7% of Africa's Gross Domestic Product (GDP), generating nearly $220 billion in economic value (Groupe Spécial Mobile Association, 2025). An important driver of this economic contribution is the growing adoption of digital commerce platforms by businesses. Previous research shows that when businesses use e-commerce applications, they grow rapidly and stimulate broader community development through positive economic cycles (Qureshi, 2023). However, these benefits are not guaranteed. When digital technologies are misused or create new vulnerabilities, they can instead reinforce negative development cycles and expose users to economic and social harm (Qureshi, 2023).

Nigeria provides a particularly important case for examining these contrasting outcomes of digital expansion. The country has experienced rapid growth in mobile device connectivity and digital commerce, while also continuing to face persistent technology-related fraud and scams that undermine trust in digital systems. These include increasingly sophisticated schemes such as fake e-commerce platforms, fraudulent job offers, and deceptive technical support operations, all of which have been linked to large financial losses among Nigerians (Microsoft, 2025). With a population exceeding 240 million, about 64% active mobile users, and smartphone penetration at 85%, Nigeria represents one of Africa's largest and most digitally-connected markets, where digital innovations are driving socio-economic development (Groupe Spécial Mobile Association, 2025; Vanguard, 2025). These characteristics make Nigeria a key setting for studying how expanding digital platforms influence governance, economic opportunity, and end-user protections.

[1]**Please cite this preprint as:** George Grispos and Sajda Qureshi, (2026), "Investigating Artificial Intelligence Digital Sovereignty in Mobile Shopping Apps: A Case Study of Nigeria", Thirty-second Americas Conference on Information Systems, Reno.

According to the United Nations (n.d.), Nigeria is classified as a medium human development country, a category frequently associated with patterns of growth linked to increased mobile phone usage (Groupe Spécial Mobile Association, 2025; Vanguard, 2025). At the same time, the 2024 Transparency International Corruption Perceptions Index ranks Nigeria 140th out of 180 countries (Transparency International, 2024), highlighting ongoing institutional challenges and underscores the need to strengthen governance mechanisms, including digital sovereignty frameworks. Despite these constraints, Nigeria's e-commerce market ranks 38th globally, is projected to reach nearly $15 billion by 2029, and over 70% of e-commerce transactions are made on a mobile device (Go-Global, 2025).

As mobile commerce expands in Nigeria, industry analysis suggests that Nigerian e-commerce platforms are increasingly integrating automated and data-driven features including personalization recommendations and automated customer interaction (GenYZ Solutions, 2023; Oloni, 2024). Given the strong dependence of Nigerian e-commerce on mobile usage, mobile applications function as primary gateways through which users engage with digital markets and are therefore well positioned to embed AI-driven functionalities that may already be shaping user experiences. However, this technological shift raises concerns about risks to Nigeria's *digital sovereignty*, the capacity of states and citizens to retain control over data, digital infrastructure, and technology, with consequences extending beyond individual users to entire societies (Floridi, 2020; Fratini et al., 2024). Digital sovereignty once referred mainly to state control over national data, but now also includes individuals' control over their personal information in relation to large technology firms, which points to broader disputes about control and authority in the digital world (Roberts, 2024). While digital sovereignty includes broader issues such as data control and infrastructure dependence, this study focuses on transparency as a measurable indicator of user awareness and control.

This paper develops the concept of digital sovereignty by investigating *how AI impacts digital sovereignty in the context of socio-economic development*? Although digital sovereignty has gained prominence, it remains an emerging and contested concept shaped by competing narratives among global actors (Fratini et al., 2024). Using an inductive analysis of AI-driven features in selected Nigeria e-commerce mobile applications, the study identifies the AI functionalities present and then evaluates how transparently providers disclose their use of AI. Transparency is then used as an indicator of digital sovereignty in Nigeria. The remainder of this paper is structured as follows. The next section outlines the theoretical background for this study, followed by a description of the methodology. The subsequent section presents the results and analysis, after which their implications for socio-economic development are discussed. The final section concludes the paper and presents ideas for future work.

## Theoretical Background

Digital sovereignty refers to the ability to exercise authority and control over digital technologies and infrastructure (Floridi, 2020; Roberts, 2024). Although originally linked to state power, recent research now views digital sovereignty as shared among governments, corporations, and individuals, who all influence how digital systems are built and governed (Tretter, 2023). Drawing on the example of COVID-19 contract tracing applications, Tretter (2023) shows how digital sovereignty operates as a network of interacting actors including nations, technology providers, and users, each exercising different forms of power that can strengthen or constrain the sovereignty of others in this ecosystem. This broader view reflects the growing role of AI, global technology platforms, and large data ecosystems in shaping economic activity, social interaction, and access to information (Roberts, 2024). Previous research has also attempted to distinguish between descriptive and normative approaches to digital sovereignty. Descriptive approaches focus on who controls digital technologies, while normative approaches consider whether that control is legitimate and accountable (Roberts, 2024). This distinction is especially relevant for AI infrastructures, where large companies control cloud services, data systems, and algorithmic platforms, giving them significant influence over digital environments, with limited oversight (Roberts, 2024).

Separately, governments are also seeking to strengthen technical autonomy and reduce reliance on external providers. Calderaro and Blumfelde (2022) show that digital sovereignty initiatives are closely linked to technological autonomy and leadership in AI capabilities. Jiang (2024) explains that national approaches to digital sovereignty vary widely among nations. Focusing on China, India and South Africa, Jiang suggests that these approaches are shaped by differences in technological capacity, autonomy, and national policy priorities. Similarly, Soulé (2024) highlights that in African contexts, digital sovereignty appears to be closely linked to issues related to infrastructure development, technology dependence, and digital capacity.

Alongside state and corporate perspectives, the literature also distinguishes between digital sovereignty at the state level with that at the individual level (Floridi, 2020; Fratini et al., 2024). At the individual level, digital sovereignty refers to a person's ability to control their personal data and digital identity, and to the extent to which technology support or constrain that control (Floridi, 2020). This perspective connects to a broader theory that suggests individuals have the capacity to share their own lives through control over their identity and decision-making (Sen, 2013). Qureshi (2022) highlights identity control as a core component of an individual's autonomy. One example of this principle is self-sovereign identity, a model in which individuals create and manage their own identities, rather than relying on central service providers (Der et al., 2017). Similarly, Stockburger et al. (2021) discuss is self-sovereign identity in terms of a user-controlled resilient identity management system that is enabled by distributed ledger technology.

Digital sovereignty appears to operate across state, corporate, and individual levels. This paper focuses digital sovereignty from the individual level by analyzing AI-driven features in selected Nigerian e-commerce mobile applications. It identifies the AI functions used and assesses how transparently providers disclose their use, treating transparency as an indicator of digital sovereignty.

# Methodology

The methodology used in this study can be described as interpretive, drawing on Klein and Myers' (1999) principle of abstraction and generalization. Data interpretation combines the forensic analysis of Android mobile applications with contextual document analysis. The first data source involves the extraction and analysis of AI-relevant artifacts from three Android e-commerce applications, Jumia, Konga, and Jiji, executed in a controlled Android environment. The second data source consists of an algorithm transparency assessment, in which publicly available information associated with each application is analyzed to identify disclosures related to AI, automated decision-making and algorithmic processing. Moreover, the study also includes contextual indicators from the 2025 United Nations (UN) Global Survey on AI and Human Development (United Nations, 2025) and the UN Human Development Index (United Nations, n.d.). Combining these sources allows AI-related forensic artifacts and transparency disclosures to be interpreted within the broader context of development environments and perceptions towards AI.

To ensure a controlled and reproduceable environment, an Android Virtual Device (AVD) was created using the Android Emulator available in Android Studio (Android Developers, n.d.-b). The AVD provides an environment that is free from personal or sensitive data and allows for replication across experiments (Ceballos Delgado et al., 2022). Since the Android security model restricts access to application directories in the Android filesystem (Android Developers, n.d.-c), the AVD was rooted using a Magisk-based method that modifies the emulator system image to enable administrator (root) access, allowing access to protected application directories in a controlled environment (Wu, n.d.). In this case, rooting is considered appropriate since the AVD contains no real user data and is consistent with previous mobile forensic practices for obtaining access to the application directories (Lessard & Kessler, 2010).

After rooting, a test Google account was created and the three Nigerian e-commerce mobile applications, Jumia, Konga, and Jiji were installed. These applications were selected following a review of industry and academic sources on the Nigerian e-commerce ecosystem (Ekhator, 2025; Olagunju et al., 2020). Cross-referencing these data sources consistently identified these applications as leading platform providers. Furthermore, Google Play Store statistics indicate over 100 million installs for Jumia, over 10 million for Jiji, and over 1 million for Konga. Each application was executed on the AVD and subjected to synthetic interactions, which includes browsing listings, modifying shopping carts, and navigating interface menus. No user accounts or credentials were created or supplied. A data extraction was then performed using the Android Debug Bridge (ADB) (Android Developers, n.d.-a), a command-line tool previously used for mobile forensic extractions (Lessard & Kessler, 2010). The application directories were identified, and ADB was used to extract each full directory to a desktop system.

The extracted application directories were then analyzed with emphasis on identifying SQLite databases, XML files, cache and temporary files, logs, and metadata, which represent common Android forensic artifacts (Grispos et al., 2013, 2015). An artifact-guided process was then applied to identify AI-related components. This involved (1) systematically screening the extracted artifacts for technical indicators such as SDK names, filenames, and logs, and (2) cross-referencing these indicators with publicly-available documentation and academic literature to determine their association with AI infrastructure and functional roles. The analysis focused specifically on (a) recommender and personalization systems, (b) conversational AI chatbots, and (c) engagement analytics platforms (Organisation for Economic Co-operation and Development, 2021). These were further classified as either locally executed or server-dependent systems based on whether user data is processed on-device or transmitted to external infrastructures. This study does not directly observe AI at runtime, but infers its presence from SDK names, filenames, logs, and other application data. The findings therefore indicate AI integration, rather than confirmed model execution.

To assess the transparency of AI usage in the three applications, a document analysis was conducted for each platform, using publicly available information. For each provider, the Google Play Store listing, linked privacy policies, and official platform websites were reviewed to identify explicit or implicit disclosures of AI, automated decision-making, personalization, or algorithmic processing. The analysis focused on algorithmic transparency, defined as "the disclosure of information about algorithms to enable monitoring, checking, criticism, or intervention by interested parties" (Diakopoulos & Koliska, 2017).

To support the interpretation of the AI-related artifacts recovered from the three applications, a contextual document analysis (Bowen, 2009) was undertaken using the UN Global Survey on AI and Human Development and the UN Human Development Index. These sources provide a framework for situating the forensic findings within Nigeria's broader development context and societal perceptions of AI. The AI and Human Development survey (United Nations, 2025) examines public interaction with AI, including familiarity, usage, trust, and perceived societal impact, while the Human Development Index (United Nations, n.d.) provides development indicators. Together, these datasets were compared with the forensic findings to identify patterns linking application behavior with socioeconomic characteristics.

# Results and Analysis

The results combine the mobile forensic findings from the three mobile applications, which were used to identify evidence of AI interactions, together with the transparency-related analysis and contextual assessment based on the 2025 U.N. Global Survey on AI and the Human Development Index.

## *Identification of AI Components*

At a high-level, the forensic analysis of the Konga, Jumia, and Jiji applications revealed a variety of artifacts reflecting user interactions and AI-driven functionality. While additional artifacts were recovered during the analysis of the applications, this study only focuses on those artifacts indicating potential links to AI functionalities. Although this additional recovered data may be valuable for a general forensic analysis, these artifacts are considered out of scope for this research.

Within the Konga application directory, artifacts were recovered from a SQLite database called `NCSmartech`. This database appears to be related to Netcore Smartech, a customer engagement and personalization platform used by Konga for AI-driven product discovery, automated campaigns, and behavioral segmentation (Netcore, n.d.). A table within this database called `event` contains timestamped activities describing product searches and interactions, navigation data, and session identifiers. Figure 1 shows an example entry from the `event` table.

{"networkMode":"Unknown","eventId":0,"attrParams":{},"lng":null,"cg":"0","timeZone":
"GMT-06:00","mid":12,"sessionId":"1770911899079","td":0,"screenOrientation":
"portrait","payload":{"Key_word":"lg tv"},"identity":"","eventTime":"1770911932446",
"eventName":"product_search","lat":null,"retry":0}

**Figure 1. Timestamped Product Search with Session Identifier**

A separate SQLite database within the Konga directory, `ChatDatabase`, contains a table `(homeDataResponse)` with JSON records that appear to contain product placement metadata (Figure 2a) and recommendation blocks (Figure 2b). Analysis of Netcore's online documentation (Netcore, n.d.) describes AI-based recommendations and engagement features, supporting the forensic interpretation that this database and its table contains user data that is part of Konga's personalization infrastructure. Additional artifacts were recovered from shared preferences XML files associated with Firebase, which is a Google platform for analytics, messaging, and remote configuration support for user engagement and personalization (Google, n.d.-a). The artifacts recovered include Remote Config files. Firebase documentation (2026) indicates that the Remote Config personalization feature applies machine learning to tailor parameter delivery for individual users. Together with the Netcore artifacts, the evidence suggests that Konga uses AI-driven personalization and engagement features within its application.

Analysis of the Jumia application folder revealed artifacts distributed across multiple files and subfolders within the parent folder. Within the SQLite database called `jumia.db`, the tables `RecentlySearch` and `RecentlyView` appear to store user search terms and viewed product identifiers, which may represent implicit feedback (Aggarwal, 2016). Although no direct model-generated recommendation data was recovered, previous research has discussed how Jumia deploys AI-based personalization architectures across its business operations (Taiwo, 2024).

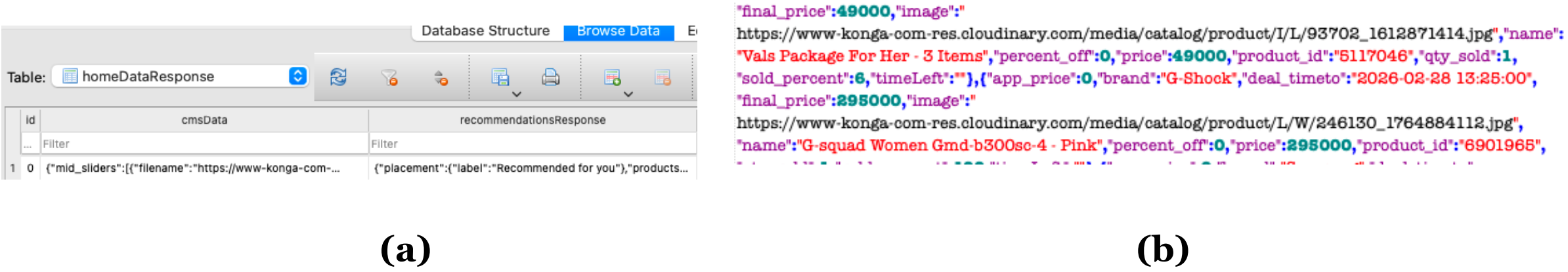

**(a)** **(b)**

**Figure 2. Product Recommendation Metadata and JSON Record Blocks**

In July 2024, Jumia announced a customized AI-powered platform incorporating conversational bots from Sprinklr (Jumia, 2024). Consistent with this announcement, the `LiveChatConfigEntity` table in `jumia.db` contains chat integration settings identifying Sprinklr as the provider (Figure 3). While chat transcripts were not recovered during the analysis of the Jumia application, Sprinklr metadata together with public deployment reports provide evidence that the application interacts with AI-enabled conversational support from Sprinklr.

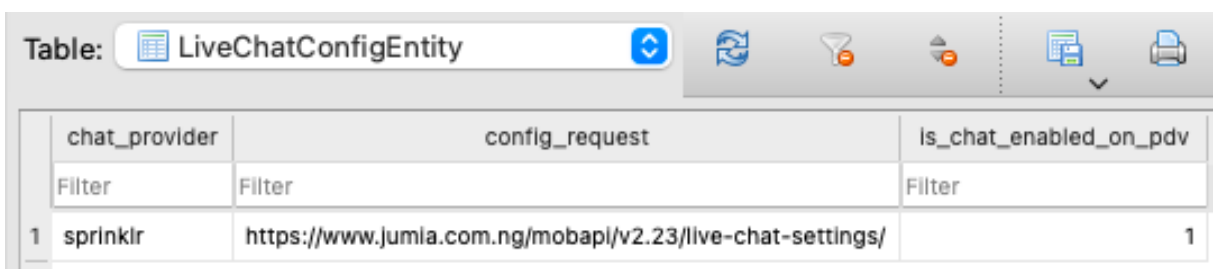

| chat_provider | config_request | is_chat_enabled_on_pdv |
|---|---|---|
| sprinklr | https://www.jumia.com.ng/mobapi/v2.23/live-chat-settings/ | 1 |

**Figure 3. Sprinklr AI Bot Integration Within Jumia**

AI-related artifacts were also recovered from files indicating integration with MoEngage, an "AI-powered marketing platform for customer engagement" (Gupta, 2025). A SQLite database, `MOEInteractions`, contains several tables including `USERATTRIBUTES`, which stores a persistent device and user identifier, and `ATTRIBUTE_CACHE`, which contains timestamped engagement attribute values. Entries in the `ATTRIBUTE_CACHE` table include recent search terms (Figure 4a.), viewed products and cart activity (Figure 4b.), and hierarchical product category metadata (Figure 4c). Taken together with the MoEngage SDK Developer Guide (MoEngage, n.d.), the recovered on-device artifacts suggest that the Jumia application stores user interactions and engagement history as part of its integration with MoEngage.

| MOE_GAID | ba8dab2c-9669-4d36-8c5c-d9f555dff7a5 | 1770912425488 |
|---|---|---|
| shop_country | NG | 1770912429156 |
| language_selection | EN | 1770912429177 |
| current_app_version | 18.3.1 | 1770912429217 |
| last_searched_term | lg tv 50 | 1770912443614 |

| last_cart_added_product_sku | LG835EL01YTHCNAFAMZ | 1770912458478 |
|---|---|---|
| last_cart_added_product_name | 55'' 4K UHD Smart TV+Netflix,YouTube APP+24 MONTHS ... | 1770912458506 |
| last_cart_added_category_key | large-screen-tv | 1770912458537 |
| last_cart_added_category_name | 50 - 65 Inches TVs | 1770912458562 |
| last_cart_added_category_level_1 | Electronics | 1770912458584 |
| last_cart_added_category_level_2 | Television & Video | 1770912458600 |
| last_cart_added_category_level_3 | Televisions | 1770912458622 |
| last_cart_added_category_level_4 | Large screen TV | 1770912458642 |

**(a)**

**(b)**

| last_viewed_category_level_1 | Home & Office | 1770912488814 |
|---|---|---|
| last_viewed_category_level_2 | Home & Kitchen | 1770912488844 |
| last_viewed_category_level_3 | Kitchen & Dining | 1770912488892 |
| last_viewed_category_level_4 | Cutlery & Knife Accessories | 1770912488918 |
| last_viewed_category_level_5 | Knife Blocks & Storage | 1770912488937 |
| last_viewed_category_level_6 | Cutlery Trays | 1770912488974 |

**(c)**

**Figure 4. Jumia and MoEngage Engagement**

The Jiji application folder includes a database (`history.db`), whose `views` table stores JSON records of viewed product listings (Figure 5), including listing IDs, product attributes, seller metadata, timestamps, and chat-availability flags. This recovery indicates that Jiji retains user interaction and messaging data commonly used by AI-enabled personalization and engagement systems (Huang & Rust, 2021).

```
"id_verify","value":"Verified ID"}],"last_seen":"1 hour ago","make_offer_range":{"max"
:375000.0,"min":175000.0},"paid_info":{"icon":true,"package_type":"premium","text"
:"Promoted"},"price_history":false,"price_obj":{"price":"₦ 250,000","value":250000,
"view":"₦ 250,000"},"published":"2 hours ago","redirect_to_similars":false,"region_id"
:164,"regions_display":["Lagos","Ojo"],"share_link":"
https://jiji.ng/ojo/tv-dvd-equipment/43-inch-lg-full-smart-17vxpXg5FjfGYcCmivLqdx6
0.html","sold_reported":false,"title":"43 Inch Lg Full Smart","title_labels":[],
"tops_count":0,"user_guid":"clpnQACyQ7Dr4veWCa4SDpzy","user_id":16287955,
"user_name":"Collins Collins Valid","user_phones":["08100710985"],"user_registered"
:"9 h","video_call":true,"X-Listing-ID":"qpRSm28LL-GE5ZTo"}
```

**Figure 5. Jiji Viewed Product Listings**

Analysis of Jiji's shared preference files and cache directory indicates integration with the Intercom messaging platform, which supports automated workflows and AI-enabled conversational services (Intercom, n.d.). The XML files `INTERCOM_SDK_PREFS.xml` and `INTERCOM_SDK_DATA.xml`, together with the JSON records in the `cache/intercomMetrics/` contain messaging identifiers, configuration metadata, and timestamped engagement events. Figure 6 shows an example of Intercom event record with timestamp, the unique event ID, SDK information, and runtime metadata confirming active messaging use.

```
{"created_at":1770912826,"id":"988b2ded-026e-4b09-8e85-
6aaeaaaf5e98","metadata":{"owner":"messenger","app_name":"Jiji.ng","app_versi
on":"6.0.1.0","app_min_sdk_version":"23","context":"no_context","sdk_version"
:"17.3.1","action":"used","place":"api","android_is_debug_build":false,"andro
id_installer_package_name":"com.android.vending","object":"set_launcher_visib
ility"},"name":"m5_metric"}
```

**Figure 6. Example of Jiji Intercom Event Record**

Further analysis of Jiji's shared preferences also suggests interaction with Google's ML Kit framework, which provides machine learning capabilities for mobile applications (Google, n.d.-b). The recovered file `com.google.mlkit.internal.xml`, includes a string named `ml_sdk_instance_id` (Figure 7), confirms that Google ML Kit components are used during the execution of the application.

```
▼<map>
   <string name="ml_sdk_instance_id">aef2bf6a-d3a7-4b6b-a625-d95a3e0fa72c</string>
 </map>
```

**Figure 7. Google ML Kit Identifier**

Across all three applications, the identified components are predominantly server-dependent systems, relying on external platforms for data processing and personalization, rather than on-device AI execution. While these findings indicate the presence of AI-related infrastructure, they do not provide direct evidence of model execution or algorithmic decision-making at runtime. However, the combination of these components and supporting documentation provide evidence of AI-based features in the applications. Similarly, stored user interaction data (e.g., search history and viewed items) do not directly show that AI is being used, but this type of data is commonly applied in personalization and recommendation systems.

## *Artificial Intelligence Transparency Analysis*

The document analysis identified limited disclosure of AI usage across the three applications. None of the Google Play Store descriptions for the applications explicitly referenced any AI-driven functionality or automated decision-making, or algorithmic systems. Consequently, prospective users receive no indication of possible AI involvement at the point of application discovery or installation. This reduces transparency at the earliest stage of user engagement. Among the three platforms, Jiji provides the clearest indication of AI use. Its privacy policy (Jiji, n.d.), explicitly states that automated tools powered by AI operate within its platform, suggestion some level of formal acknowledgement of algorithmic processing. In contrast, Jumia's privacy policy (Jumia, n.d.) outlines the use of customer data for personalization, behavioral analysis, and service improvement. While these practices imply the presence of recommendation systems or profiling mechanisms, the document does not explicitly reference AI or machine-learning technologies, leaving the technological basis of such processing only indirectly communicated.

Konga exhibits the lowest observable level of transparency. Neither its Google Play Store listing, nor its privacy policy documentation (Konga, n.d.) contain explicit references to AI, automated decision-making or algorithmic processing. The absence of such disclosure makes it difficult for users to recognize the presence of AI systems, particularly given earlier forensic findings indicating that Konga deploys AI-driven personalization and engagement tools within its application. Overall, the analysis of the platforms' level of disclosure regarding AI usage in user-accessible documentation is limited, and when present, tends to appear only within detailed legal privacy policies, rather than prominently in user-facing materials.

### *Socio-economic Development Analysis*

The analysis of the survey and the Human Development Index data reveal several interesting observations related to AI adoption, awareness and trust in Nigeria. The data indicates that AI adoption in Nigeria can be characterized as moderate awareness, uneven interaction frequency, but a strong dependence on consumer digital platforms. While nearly 40% of the respondents indicated knowing only "a little" about AI, nearly 22% reported no knowledge at all, with only 13% indicating a high level of familiarity with AI.

Direct interactions with AI systems and services appear to be uneven. Nearly 41% of the respondents indicated that they had never knowingly interacted with AI, compared to just over 25% who answered interacting with AI more than once per week. In terms of reported AI usage, the respondents indicated that they use AI primarily in consumer-facing domains, including social media (39%), education platforms (26%), and entertainment services (21%). Interestingly, AI interactions are substantially lower within government services (7%) and healthcare (12%). Regardless, the respondents indicated that they have high expectations for future AI use, with nearly 57% indicating that AI is very likely to be used in education and 58% suggesting entertainment or social media applications. From a trust perspective, the results indicate a low level of public trust in AI systems, with less than half of the respondents (48%) indicating either strong or moderate confidence, while nearly 52% expressed limited or no confidence in AI. These findings indicate that the Nigerian public's trust in AI is still cautious, and that familiarity with AI does not necessarily lead to confidence in the reliability, transparency, or regulatory oversight of these systems. This is consistent with previous studies (Glikson & Woolley, 2020; Siau & Wang, 2018) that show that AI adoption depends heavily on perceived reliability, explainability, and governance protections.

## AI and Digital Sovereignty in Socio-Economic Development

The concepts emerging from this research suggest that AI affects individual digital sovereignty through limited user awareness, strong dependence on consumer digital platforms, uneven direct interactions with AI systems, and relatively low public trust in AI technologies. The widespread presence of AI in mobile applications, often with limited user knowledge or consent, highlights the growing importance of digital sovereignty. Although AI-enabled features and related artifacts are common in the evaluated mobile applications, the document analysis identified limited disclosure of AI usage across the three applications.

These findings support analytical generalization (Klein & Myers, 1999), rather than statistical generalization. In particular, the study generalizes the relationship between the use of AI, limited transparency, and reduced individual digital sovereignty identified in the analyzed applications. This type of generalization is most relevant to mobile-first digital economies that rely heavily on digital platforms, have developing regulatory frameworks, and show moderate levels of AI awareness, such as Nigeria and similar emerging markets. The findings are therefore not intended to apply to all e-commerce contexts but instead aim to support theory-building on how AI-enabled platforms may influence individual digital sovereignty in similar socio-technical environments. Drawing on Walsham (1993), such generalization may support concept development, theory generation, and the identification of broader impacts.

Several implications for socio-economic development follow. AI features such as recommender systems are not inherently harmful; concerns arise when their operation lacks transparency. Many of these risks exist without AI, but AI amplifies them through automation, scale, and predictive capabilities. First, technological trust is critical (Cunha et al., 2021). AI systems collect personal information through mobile applications and feed this data into large language models that interact with users, raising concerns about data control and consent. Second, from a technical perspective (Cunha et al., 2021), AI relies on user input to support conversational agents that recommend products and services, increasing the scale and efficiency of data sharing. Third, discriminative models can classify existing data through supervised learning and predict how user data should be categorized, often without transparency regarding how the data is used.

Such predictive profiling can reduce trust, especially when models generate highly accurate behavioral predictions. AI systems using predictive analytics may aggregate data from multiple sources to construct detailed user profiles containing thousands of data points (Grispos et al., 2014). Zuboff (2015) describes this process as surveillance capitalism, in which data-driven systems not only predict consumer behavior, but also shape consumer actions. Although AI models are designed to predict consumer behavior, such systems can also enable large-scale monitoring and modification of human actions, raising concerns about the potential for AI-driven social control. Similarly, strengthening individual digital sovereignty could also help foster positive cycles of development through digital innovation (Qureshi, 2023).

## Conclusions and Future Work

This research advances the understanding of individual digital sovereignty by providing empirical evidence that links AI transparency with user trust and individuals' control over their personal data and digital devices. As AI becomes increasingly integrated into mobile applications, safeguarding individual digital sovereignty is growing in importance. Consistent with this concern, the document analysis findings align with the forensic evidence recovered from the examination of the three Nigerian e-commerce applications. The recovery of configuration files, user engagement databases, recommendation data, messaging telemetry, and machine-learning SDK components indicates the use of external personalization engines, automated customer interaction system and behavioral targeting infrastructure by the three providers. Collectively, these results suggest that Nigerian users frequently encounter AI within mobile applications such as Konga, Jumia, and Jiji, often without explicit awareness of interacting with these technologies.

This study focuses on a single country, which limits how broadly the findings can be generalized. However, Nigeria serves as an important example of a mobile-first digital economy with rapid platform growth and developing governance structures. The findings may therefore be relevant to other emerging markets with similar characteristics, although differences in regulation and levels of AI awareness should be considered. Future research plans to explore technical and architectural approaches that will provide users with clearer transparency, stronger consent mechanisms, and improve content controls in mobile e-commerce applications. Future research could also complement the forensic analysis with behavioral or black-box evaluation techniques, such as controlled interaction scenarios, to observe how system outputs (e.g., recommendations) adapt over time under different conditions. Future work can also extend the analysis presented in this paper to include a broader range of e-commerce applications across Africa, treating such platforms as important units of analysis for assessing digital sovereignty. The inductive mixed methods approach used in this research may also be applied to larger samples of widely used mobile applications to further examine how AI adoption influences user awareness, trust, and control in evolving digital markets.